\documentclass[10pt]{article}
\usepackage{amssymb,color}
\usepackage{amsmath,graphicx,enumerate,textcomp}
\usepackage[abbrev]{amsrefs}
\usepackage[a4paper,top=1.in, bottom=1.5in, left=.8in, right=.8in, headsep=0.in, centering]{geometry}
\usepackage{amssymb,amsfonts,amsthm}
\usepackage{epsfig}
\usepackage{lipsum}
\usepackage{graphicx}

\numberwithin{equation}{section}

\newcommand{\eps}{\varepsilon}

\title{On dynamics of gasless combustion in slowly varying periodic media: periodic fronts, their stability and propagation-extinction-diffusion-reignition pattern.
}

\author{Amanda Matson
\thanks{Department of Mathematical Sciences, Kent State University, Kent,
OH 44242, USA. E-mail:{\tt amatson2@kent.edu}}
\and Leonid Kagan
  \thanks{School of Mathematical Sciences, Tel Aviv University,
Tel Aviv 69978, Israel. E-mail: {\tt kaganleo@tauex.tau.ac.il }}
\and Claude-Michel Brauner
\thanks{Institut de Math\'ematiques de Bordeaux UMR CNRS 5251, Universit\'e de Bordeaux, 33405 Talence, France. E-mail: {\tt claude-michel.brauner@u-bordeaux.fr}}
\and Gregory Sivashinsky
 \thanks{School of Mathematical Sciences, Tel Aviv University,
Tel Aviv 69978, Israel.  E-mail:{\tt grishas@tauex.tau.ac.il}}
\and Peter V. Gordon
\thanks{Department of Mathematical Sciences, Kent State University, Kent,
OH 44242, USA.  E-mail: {\tt gordon@math.kent.edu}}
}

\begin{document}

\maketitle
\begin{abstract}
In this paper we consider a classical model of gasless combustion in a one dimensional formulation under the assumption of ignition temperature kinetics. We study the propagation of  flame fronts
in this model when the initial distribution of the solid fuel is a spatially  periodic function that varies on a large scale. It is shown that in certain parametric regimes the model
supports periodic traveling fronts. An accurate asymptotic formula for the velocity of the flame front is derived and studied. The stability of periodic fronts is also explored, and a critical condition in terms of parameters of the problem is derived.  It is also shown that the instability of periodic fronts, in a certain parametric regimes,
results in a propagation-extinction-diffusion-reignition pattern which is studied numerically.

\end{abstract}

\medskip


\medskip
\section{Introduction}\label{s:intro}

The present study is concerned with combustion waves in gasless systems in which the final products, as well as initial reactants, are in the solid phase.  The premixed reactants are typically powdered mixtures of metals with a nonmetal oxide.  When ignited, these systems support a self-propagating deflagration.  The combustion process results in a solid product commonly referred to as self-propagating high temperature synthesis (SHS) due to its application to the high-temperature synthesis of new refracting materials \cite{Gorosh2022, Mer2003,Mer1997}. 
A substantial part of theoretical and experimental studies of gasless combustion are targeted to understand  the dynamics of flame fronts propagating toward unburned spatially  homogeneous solid fuel. 
It is well established by now that such fronts commonly suffer from various instabilities.
Apparently the  first experimental study  reporting unstable oscillatory flames in solid samples  is 
\cite{bel50} and dates back to 1950's. 
Later experimental studies revealed  more complicated regimes of propagation such as spinning
fronts \cite{mer73}, other regimes were discovered in \cite{ex1,ex2}. The theoretical exploration of  pulsating fronts in a one dimensional configuration was performed in \cite{matsiv78}. The spinning regime in cylindrical
samples was studied in \cite{siv81}, spiral waves in planer geometry  were discussed in \cite{volmat}. Later theoretical studies (see e.g  \cite{BM90})  showed existence of non-trivial
chaotic patterns of propagations. Reviews of some of these results can be found in \cite{Marg1991,V3}.
Systematic description of experimentally observed modes of propagations is given in \cite{Ivl3}.
Earlier studies of instabilities were based on a standard single temperature constant density approximation adiabatic model with a first order reaction rate and Arrhenius temperature kinetics. Some additional effects such as  the impact of temperature kinetics and the reaction order \cite{vol2,gol}, the presence of a  heat loss  \cite{kur}  and other factors were later explored. 
Let us note that the majority of these results were obtained by formal linear stability analysis 
of flame fronts, via numerical simulations or combination of thereof.
Rigorous analysis of nonlinear stability of gasless flame fronts poses  a very non-trivial
mathematical challenge as the absence of material diffusivity leads to the analysis of
highly degenerate systems. Apparently the most advanced results in this direction, so far,
were obtained in \cite{ghaz10}.

While the understanding of propagation regimes of  flame fronts moving towards homogeneous unburned solid reactant is of major  importance, it is also of interest to consider a situation
in which the initial profile of an unreacted solid varies, particularly in a periodic manner.
Indeed, in real systems it is quite common to have a certain level of inhomogeneity in
solid reactants which, in extreme case of solids consisting of separated solid grains,   leads to emergence of  discrete reaction waves \cite{Muk2008}.
Unlike homogenous case, the theoretical explorations of flame fronts in the solids of varying concentration are more limited. It is apparent, however, that the presence of variations in solid fuel concentration
results in new effects such as, for example, frequency locking \cite{mdk, mdk2} and hump effect
\cite{CMB2}.

In this paper we are interested in the analysis of flame fronts propagating
toward unburned solid fuel whose concentration is varying on a large scale (scales much larger than the characteristic width of the flame front). We show that the
periodicity in fuel concentration leads to the emergence of periodic flame fronts.
We give a comprehensive asymptotic description of such fronts  and study their stability. We show
that the instability of periodic fronts, in certain parametric regimes, leads to a very interesting propagation-extinction-diffusion-reignition pattern.

The paper is organized as follows. In section \ref{s:form} we give a mathematical formulation of the
problem. In section \ref{s:periodic} we derive an asymptotic expression for the velocity of 
periodic traveling fronts and compare results with numerical simulations.
 In section \ref{s:stab} we discuss stability of periodic traveling fronts. In section \ref{s:pattern}
 we discuss a  propagation-extinction-diffusion-reignition  regime which appears as a result of flame front
 instability.

\section{Formulation}\label{s:form}

The current study is based on  the classical model of gasless combustion. This model, in one dimensional formulation, reads:
\begin{eqnarray}\label{eq:i1}
\left\{
\begin{array}{lll}
\Theta_t=\Theta_{xx} +W(\Theta,\Phi),  \\
\Phi_t=-W(\Theta,\Phi),
\end{array}
\right.
\end{eqnarray}
where $\Theta$ and $\Phi$ are appropriately scaled  temperature and concentration of deficient solid reactant,  $x\in\mathbb{R},$ $t>0$ are spatiotemporal coordinates  and $W(\Theta,\Phi)$ is the reaction rate.
In this work we will assume that the reaction rate has a first order reaction and ignition type temperature kinetics. Namely,
\begin{eqnarray} \label{eq:i2}
W(\Theta,\Phi):=\left\{
\begin{array}{ll}
A\Phi & \Theta\ge \Theta_i,\\
0 & \mbox{otherwise},
\end{array}
\right.
\end{eqnarray}
where $\Theta_i\in (0,1)$ is an ignition temperature and 
\begin{eqnarray}\label{eq:i3}
A=\frac{\Theta_i}{1-\Theta_i}
\end{eqnarray}
is a scaling factor. The scaling factor is chosen in such a way that the velocity of the flame
front in a spatially homogenous concentration field is unity.
 A comprehensive study
 of such flame fronts and their stability in the presence of material diffusion is given in \cite{BGKS,CMB1}. A fully non-linear reduced model for diffusive instabilities of such fronts
 was derived and analyzed  in \cite{dima}.

\begin{figure}[h!]
\centering \includegraphics[width=4.5in]{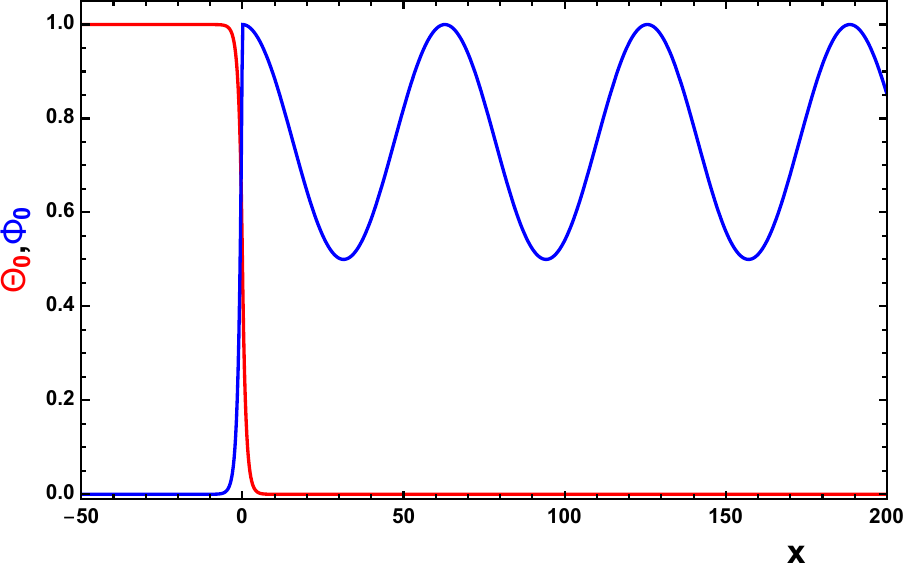}
\caption{Profile of initial conditions for the temperature $\Theta_0$ (red) and concentration of the deficient solid reactant $\Phi_0$ (blue)    with $\Theta_i=1/2$, $a=0.25$ and $\eps=0.1$ }
\label{f:1} 
\end{figure}

The goal of this paper is to analyze flame propagation regimes for model \eqref{eq:i1}-\eqref{eq:i2} 
with  front like initial conditions for the temperature and spatially periodic 
slowly varying conditions for the concentration of the deficient solid reactant. Specifically, we study problem \eqref{eq:i1}-\eqref{eq:i2}  complemented with boundary like conditions for the temperature $\Theta$ in far fields
\begin{eqnarray}\label{eq:i4}
&&\Theta(x,t)\to  0,  \quad \mbox{as} \quad x\to\infty ,\nonumber\\
&& \Theta_x(x,t)\to  0,  \quad \mbox{as} \quad x\to-\infty ,
\end{eqnarray}
and front like initial conditions:
\begin{eqnarray}\label{eq:i5}
\Theta(x,t=0)=\Theta_0(x),
\end{eqnarray}
where $\Theta_0$ is a smooth monotone decreasing function that approaches zero and one as $x\to \pm\infty$ sufficiently fast. In addition we assume that $\Theta_0(0)=\Theta_i.$
The initial concentration of solid deficient  reactant is prescribed as follows:
\begin{eqnarray}\label{eq:i6}
\Phi(x,t=0)=\Phi_0(x)
\end{eqnarray}
We set the  following assumption on the initial concentration field $\Phi_0$. The function $\Phi_0$ is an increasing function on $x\in(-\infty,0)$ that approaches zero as $x\to-\infty$ sufficiently fast 
and $\Phi_0(0)=1$. For $x\in (0,\infty)$ we set $\Phi_0(x)=\Psi_{\eps}(x)=\Psi(\eps x)$. Here and below $0<\Psi<1$ is a prescribed continuous $2\pi$ periodic function on $[0,\infty)$  satisfying $\Psi(0)=1$ and $0<\eps \ll 1 $.
That is $\Psi_{\eps}$ is a slowly varying periodic function with characteristic scale $1/\eps$.
 In numerical simulations and formal asymptotic constructions presented in this paper we set:
\begin{eqnarray}\label{eq:i7}
\Psi(x)=1-a(1-\cos(x)),
\end{eqnarray} 
with $0<a<1/2$.
Representative initial conditions for temperature and concentration of the deficient reactant are depicted in Figure \ref{f:1}.

To characterize the propagation phenomena for problem \eqref{eq:i1}-\eqref{eq:i6} it is convenient to introduce the notion of an {\it ignition interface} which is a leftmost position
in the flame front such that the maximal temperature to the right of this position is below the ignition temperature. Namely, we define the position of the ignition interface $q(t)$ as 
follows:
\begin{eqnarray}\label{eq:i8}
 q(t)=\min \{x ~ \mbox{such that} \quad  \Theta(y,t)<\Theta_i \quad  \mbox{for} \quad y>x\}.
 \end{eqnarray}
 The velocity of the ignition interface is defined as the derivative of this position, that is $d q(t)/dt$.
 It is easy to verify that when crossing the ignition interface the temperature and its spatial first derivative remain continuous:
 \begin{eqnarray}
 \Theta(q(t),t)=\Theta_i, \quad [\Theta(q(t),t)]=0, \quad [\Theta_x(q(t),t)]=0,
 \end{eqnarray} 
where $[\cdot ]$ stands for the jump of the quantity.
We note that $q(t)$ is a well defined quantity even in the absence of the reaction.  

As a first step of our study we show (on the formal level)  that problem  \eqref{eq:i1}-\eqref{eq:i6} 
admits periodic traveling fronts solutions. We construct these solutions and derive a formula
for the speed of propagation in the next section.

\section{ Periodic traveling fronts.}\label{s:periodic}

In this section we construct an approximation of periodic fronts that emerge as a long term limit of solutions for problem \eqref{eq:i1}-\eqref{eq:i6} in certain parametric regimes. 

It is well known that
the dynamics of flame fronts for model \eqref{eq:i1}-\eqref{eq:i6}  are predominately defined by their behavior in the vicinity of the ignition temperature rather than far fields. In view of this
observation and the fact that the fuel concentration field varies only on a large scale $1/\eps,$  one may expect that in the first approximation the dynamics of flame fronts
can be captured by considering the problem  \eqref{eq:i1}-\eqref{eq:i6} on the intermediate scale $1/\sqrt{\eps}$. On spacial scale $1/\sqrt{\eps}$ the  initial fuel concentration field can
be viewed as a locally prescribed constant. Hence, introducing a traveling front ansatz  $(\Theta,\Phi)(x,t)=(\theta,\phi)(\xi)$ with $\xi=x-{\eps^{-1}}{q(\tau)},$ where   $\tau=\eps t$ and $q(\tau)$
are slow time and  the position of the ignition interface respectively, and 
substituting this ansatz into \eqref{eq:i1} for sufficiently small $\eps$  one ends up with the following auxiliary problem:
\begin{eqnarray}\label{eq:fr1}
\left\{
\begin{array}{lll}
\theta^{\prime\prime}+V \theta^{\prime}=0, & V\phi^{\prime}=0, & \xi>0\\
\theta^{\prime\prime}+V \theta^{\prime}+A\phi=0, & V\phi^{\prime}-A\phi=0, & \xi<0,
\end{array}
\right.
\end{eqnarray}
where $V=d q(\tau)/d\tau$ is the velocity of the ignition interface and prime stands for the derivative with respect to $\xi$.
This problem is complemented with the continuity of the solution when crossing the ignition  interface which is set to be located at $\xi=0$
\begin{eqnarray}\label{eq:fr2}
\theta=\Theta_i, \quad [\theta]=[\theta^{\prime}]=[\phi]=0, \quad \mbox{at} \quad \xi=0 ,
\end{eqnarray}
and far field conditions
\begin{eqnarray}\label{eq:fr3}
\left\{
\begin{array}{lll}
\theta\to 0, & \phi\to \Sigma & \mbox{as} \quad \xi \to \infty,\\
\theta^{\prime}\to 0, & \phi\to 0, & \mbox{as} \quad \xi \to -\infty,
\end{array}
\right.
\end{eqnarray}
where $\Sigma$ is the local concentration of solid fuel.

\begin{figure}[h!]
\centering \includegraphics[width=5in]{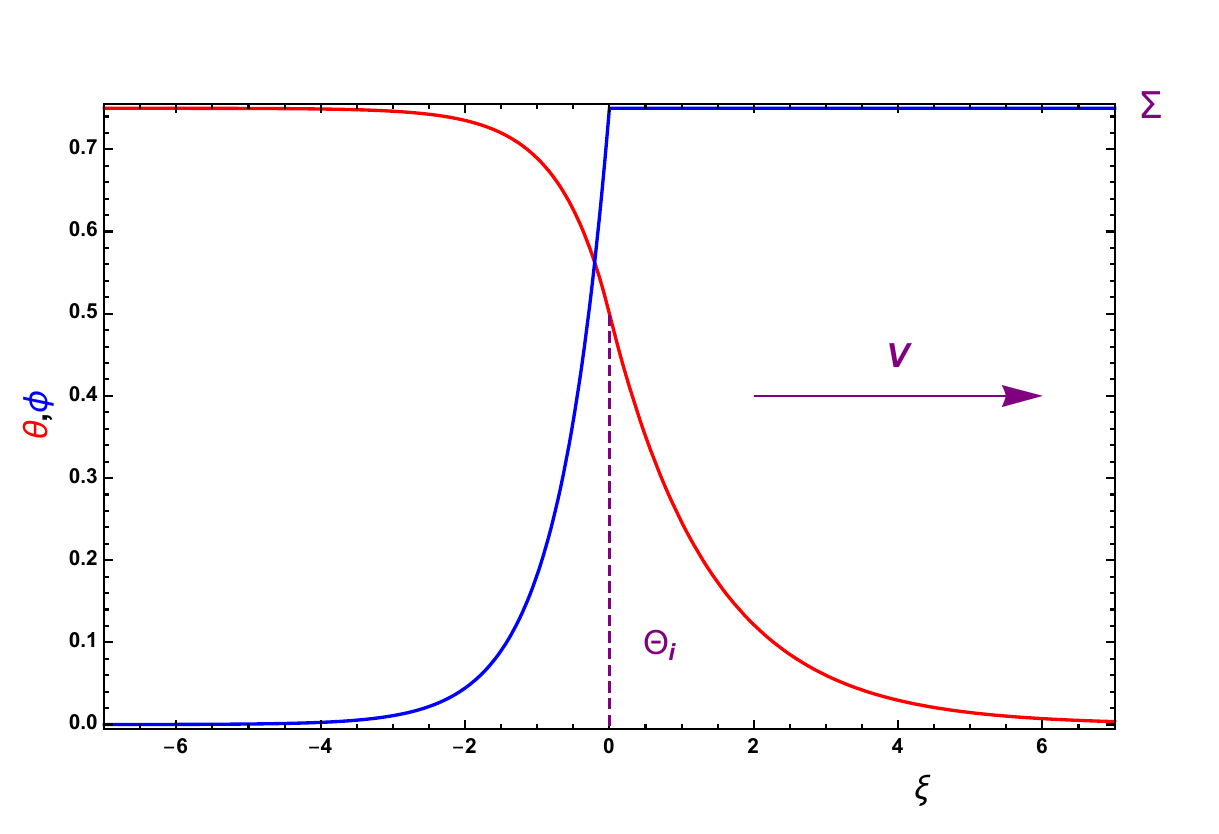}
\caption{Traveling front profile given by \eqref{eq:fr4}, \eqref{eq:fr5} with $\Theta_i=1/2$ and $\Sigma=3/4$.  }
\label{f:2} 
\end{figure}

One can easily verify that for arbitrary $0< \Theta_i<\Sigma \le 1$ this problem admits a unique solution that reads:
\begin{eqnarray}\label{eq:fr4}
\theta(\xi)=\left\{
\begin{array}{ll}
\Theta_i \exp\left(-\sqrt{\frac{\Sigma-\Theta_i}{1-\Theta_i} }\xi\right), & \xi>0,\\
\Sigma-\left(\Sigma-\Theta_i \right)\exp\left(\frac{\Theta_i}{\sqrt{(1-\Theta_i)(\Sigma-\Theta_i)} }\xi\right), & \xi\le 0,
\end{array}
\right.
\end{eqnarray}
\begin{eqnarray}\label{eq:fr5}
\phi(\xi)=\left\{
\begin{array}{ll}
\Sigma, & \xi>0,\\
\Sigma\exp\left(\frac{\Theta_i}{\sqrt{(1-\Theta_i)(\Sigma-\Theta_i)} }\xi\right), & \xi\le 0,
\end{array}
\right.
\end{eqnarray}
\begin{eqnarray}\label{eq:fr6}
V=V(\Sigma,\Theta_i)=\sqrt{\frac{\Sigma-\Theta_i}{1-\Theta_i}}.
\end{eqnarray}
A typical profile of this traveling front is depicted in Figure \ref{f:2}.
In the vicinity of characteristic size $1/\sqrt{\eps}$ of the ignition interface located at the point $ x=\eps^{-1} q(\tau)$ the shape of the flame front is expected to be close to the one given by $\eqref{eq:fr4}$ and the position
of this interface is defined by the following first order initial value problem
\begin{eqnarray}\label{eq:fr7}
\frac{d}{d\tau} q(\tau)=\sqrt{\frac{\Psi(q(\tau))-\Theta_i}{1-\Theta_i}}, \quad  t>0, \quad q(0)=0.
\end{eqnarray}
For any given function $\Psi$ such that $\Sigma_{min}:=\min_{s\ge 0} \Psi(s)\ge \Theta_i$ the solution of problem \eqref{eq:fr6} can be obtained (at least in the implicit form) by separation of variables.
In particular, when $\Psi$ is given by \eqref{eq:i7} we have the following, closed form, expressions for the position of the ignition interface
\begin{eqnarray}\label{eq:fr8}
q(\tau )=2 {\rm am}\left(\frac{\tau}{2},\lambda \right),
\end{eqnarray}
and 
\begin{eqnarray}\label{eq:fr9}
\frac{d}{d\tau} q(\tau )= {\rm dn}\left(\frac{\tau}{2},\lambda\right),
\end{eqnarray}
for its velocity. Where,
\begin{eqnarray}
\lambda:=\frac{1-\Sigma_{min}}{1-\Theta_i}=\frac{2a}{1-\Theta_i},
\end{eqnarray}
and  ${\rm am}\left(\frac{\tau}{2},\lambda \right),{\rm dn}\left(\frac{\tau}{2},\lambda \right),$ are Jacobi amplitude and Jacobi elliptic functions respectively \cite{AS}.
 \begin{figure}[h!]
\centering \includegraphics[width=4.7in]{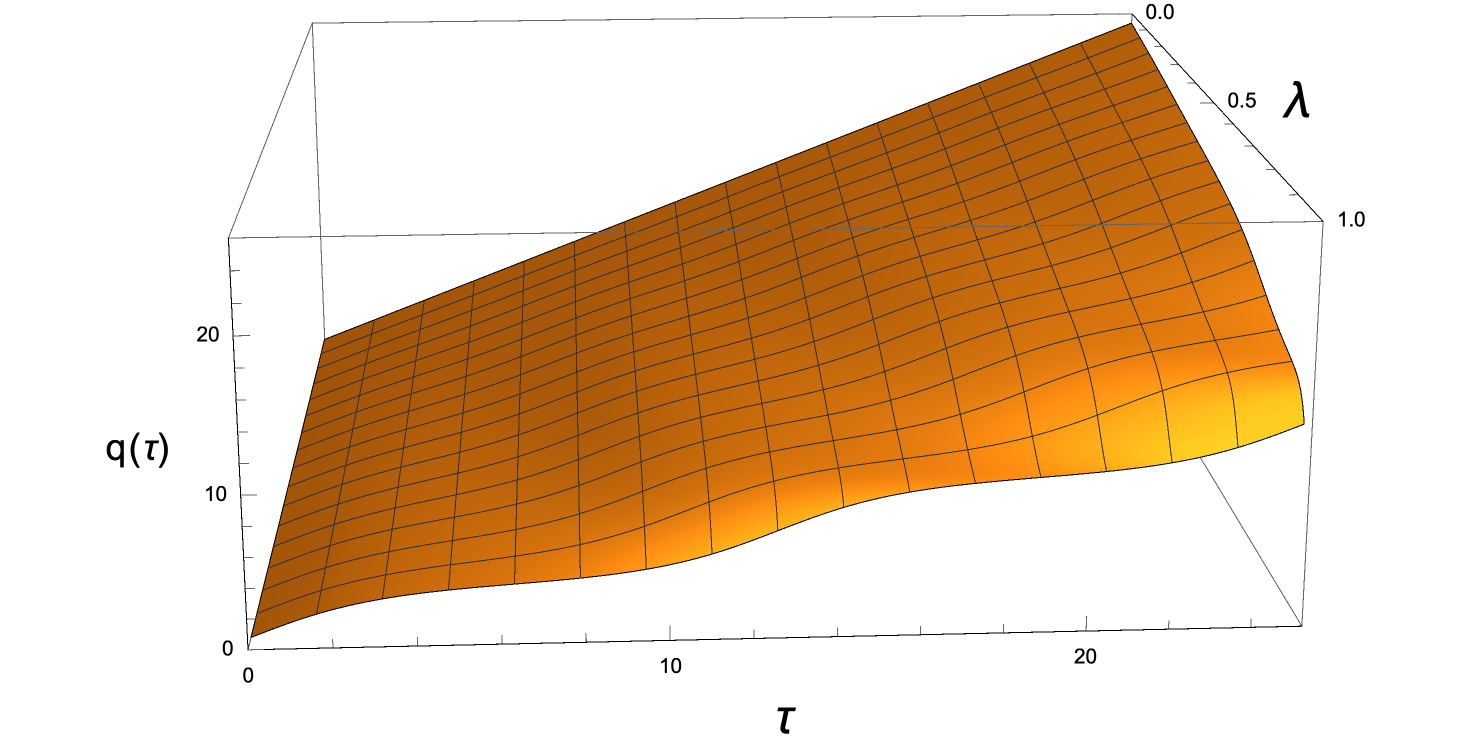}\\
\centering \includegraphics[width=4.5in]{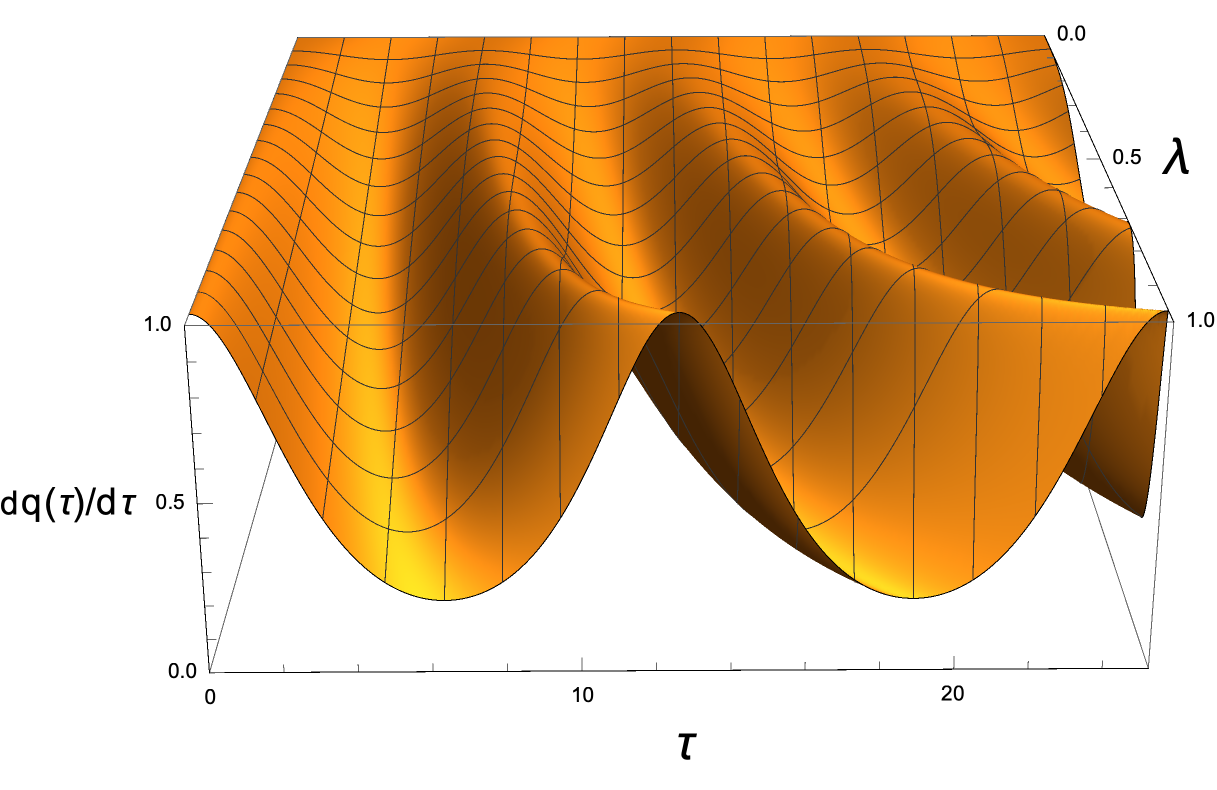}
\caption{The position $q(\tau)$ and velocity $d q(\tau)/d\tau$  of the ignition interface as a function of slow time $\tau$ and parameter $\lambda$.   }
\label{f:3} 
\end{figure}
 Hence the dynamics of the ignition front is fully determined by the specific value of the parameter $\lambda\in[0,1)$. For fixed ignition temperature $\Theta_i$ this parameter characterizes the amplitude of oscillations in the initial
 data for the deficient reactant and increases from zero to one as the amplitude of oscillations increases. We note that $\lambda\to 0$ corresponds to the case when $\Sigma_{min}=1$
 (that is the absence of the oscillations in the initial data for the deficient reactant) in which case the ignition front propagates with the constant velocity one.
 The case $\lambda\to1$ corresponds to a situation when oscillations in the initial value of the deficient reactant are maximal. In the latter  case the front is stagnant, and its velocity becomes zero when the front reaches the position with minimal concentration of solid fuel.  The  ignition front does not exist
 when $\lambda>1$.
 The profiles for the position and velocity of the ignition interface as a function of slow time $\tau$ and parameter $\lambda$ are depicted in  Figure \ref{f:3}.

\begin{figure}[h!]
\centering \includegraphics[width=4.in]{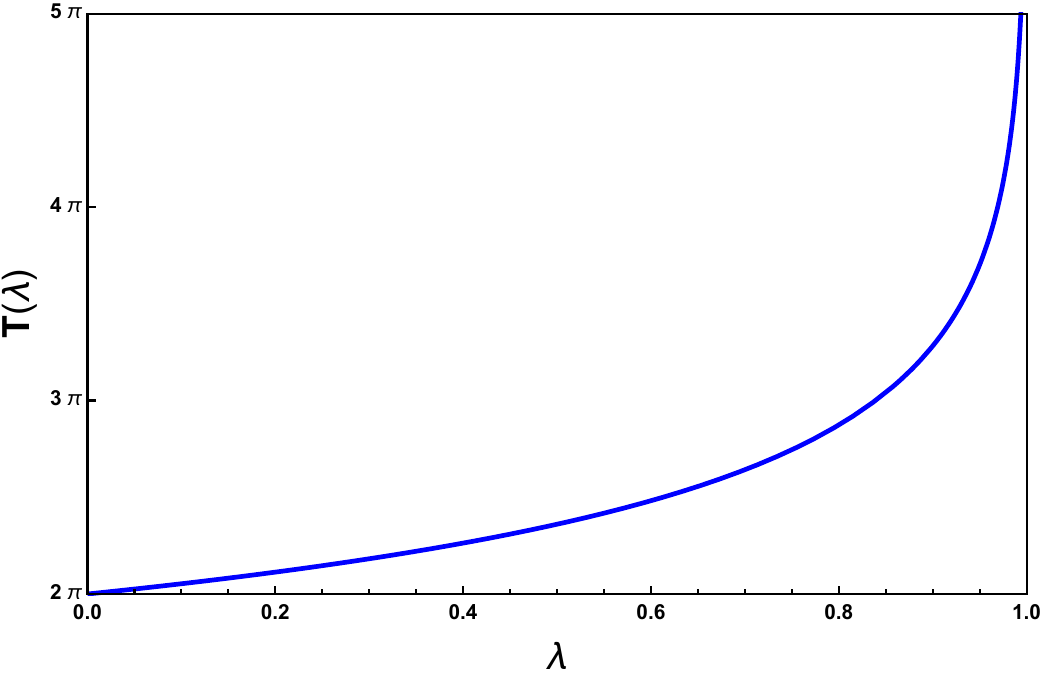}
\caption{Period of oscillation of the velocity profile as a function of the parameter $\lambda$.   }
\label{f:4} 
\end{figure}

\begin{figure}[h!]
\centering \includegraphics[width=3.2in]{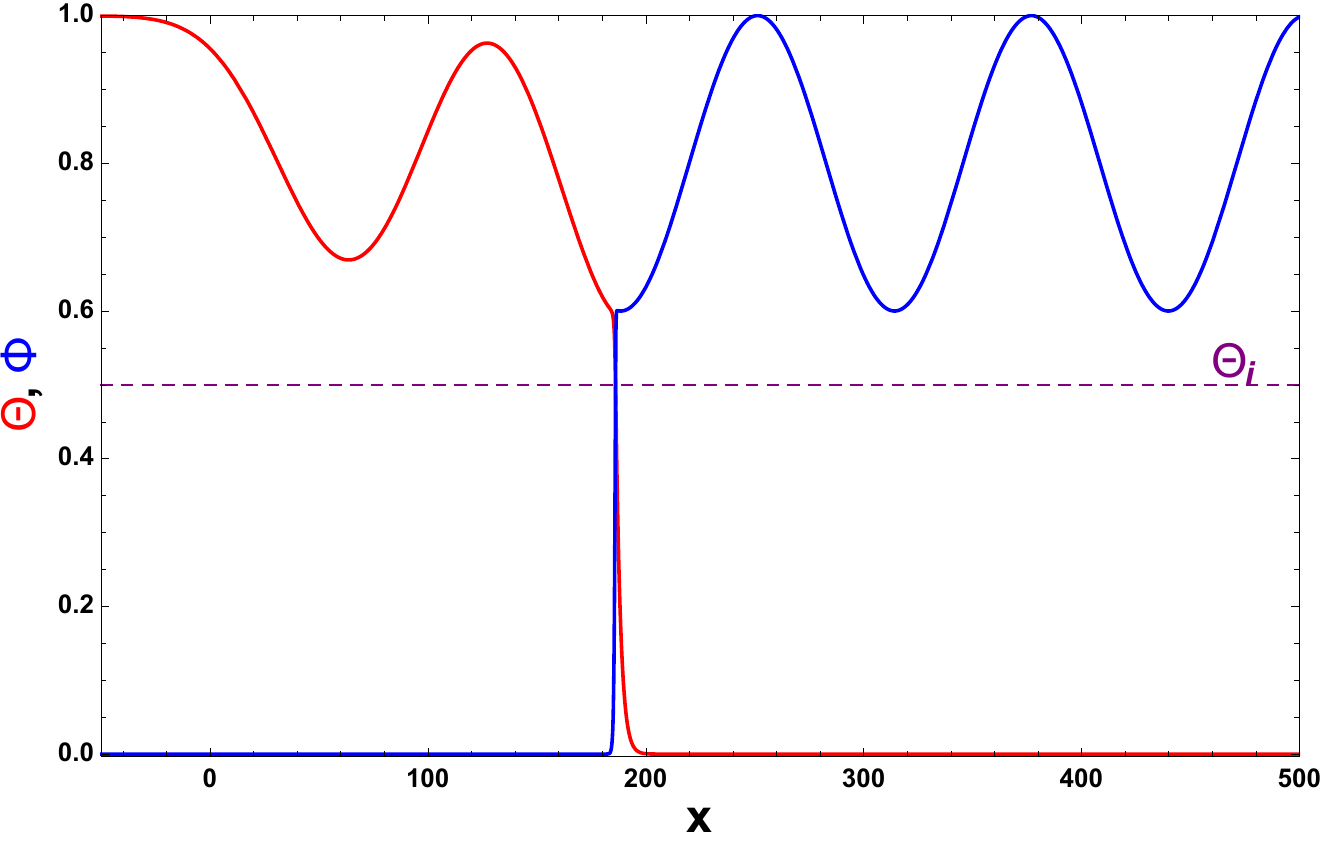}
\centering \includegraphics[width=3.2in]{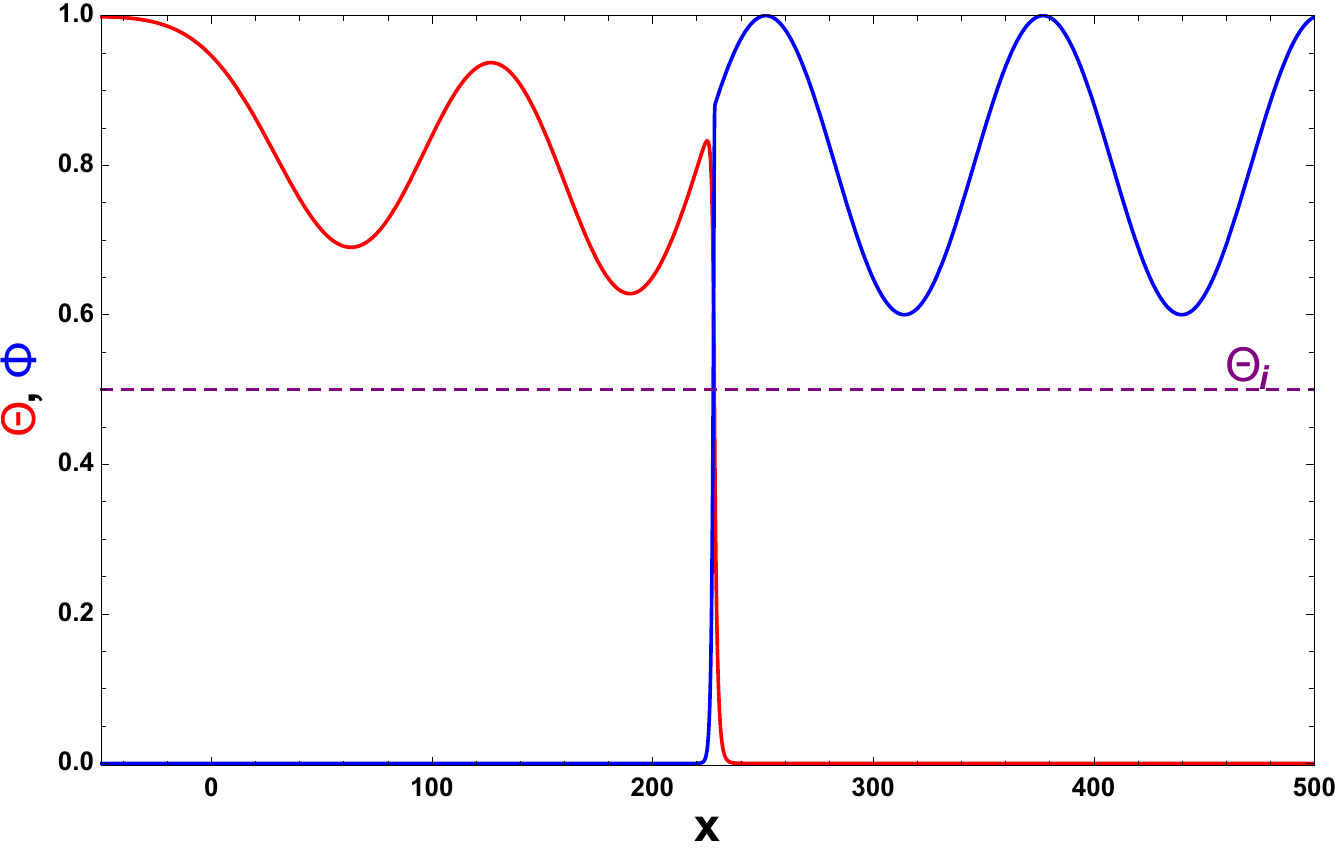}\\
\centering \includegraphics[width=3.2in]{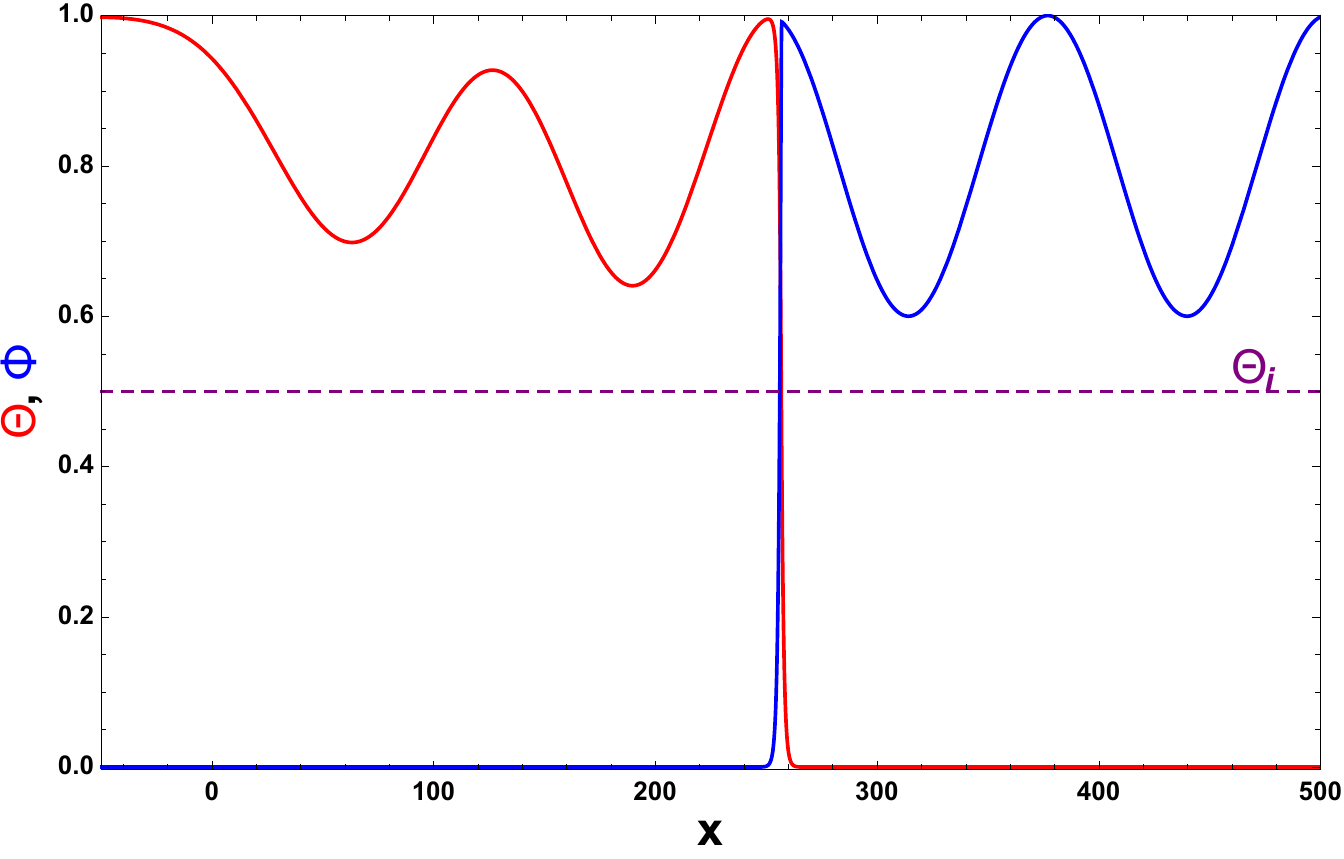}\\
\centering \includegraphics[width=3.2in]{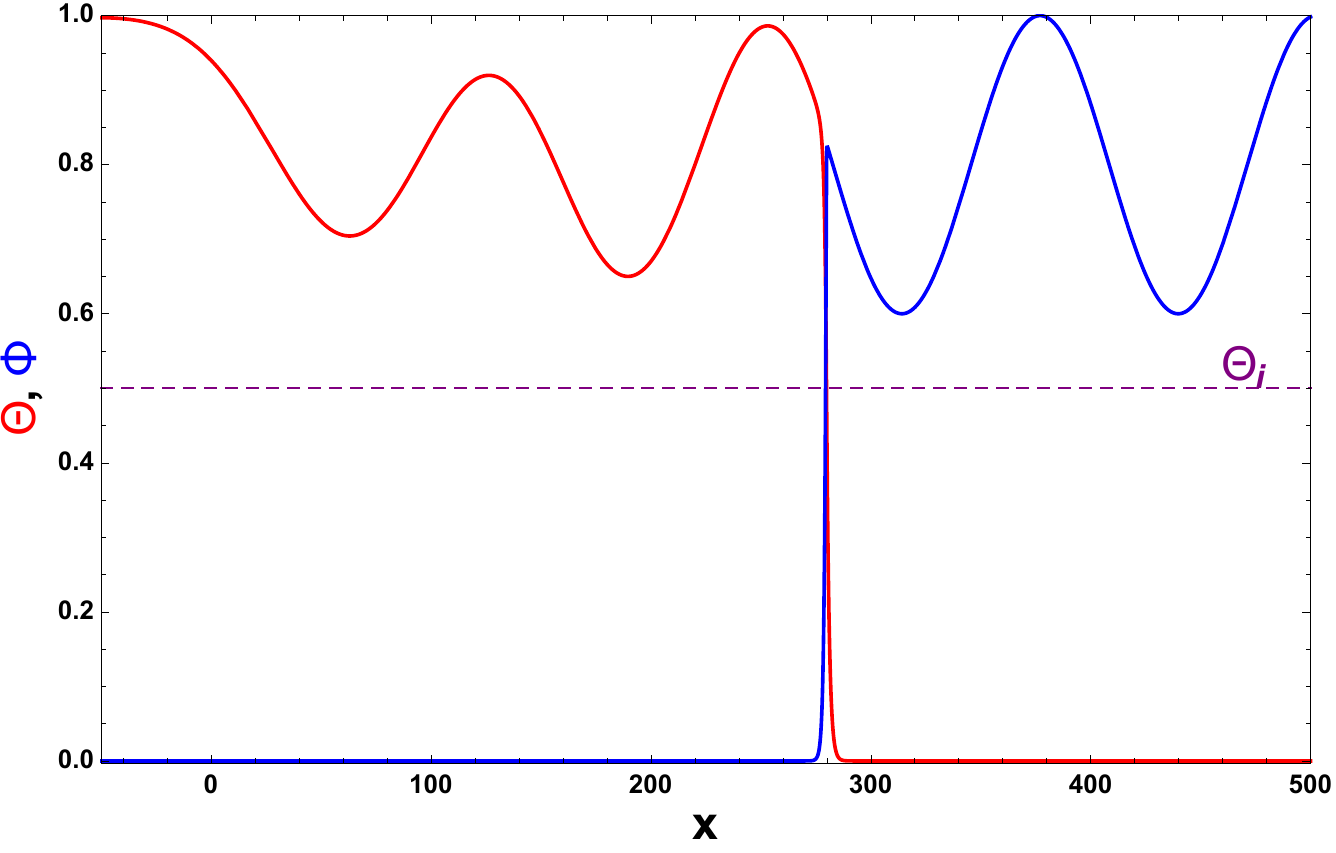}
\centering \includegraphics[width=3.2in]{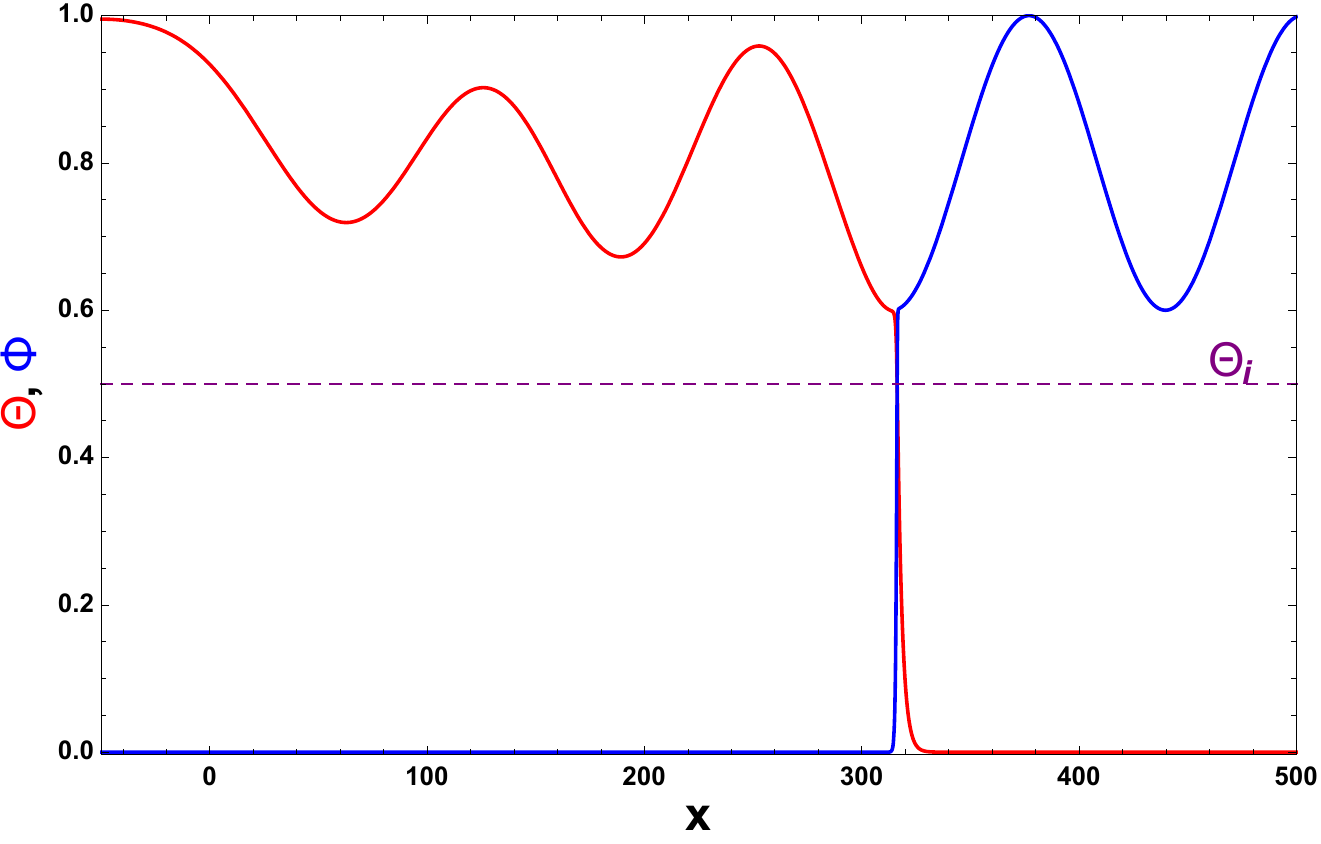}\\
\caption{Snapshots of periodic traveling front profiles for problem \eqref{eq:i1}--\eqref{eq:i6} with $\Theta_i=1/2,$ $\Sigma_{min}=0.6~(a=0.2)$ and $\eps=0.05$.
The red line represents the temperature profile and the blue line represents the profile of the deficient reactant. Purple dashed line indicates value of the ignition temperature. }
\label{f:5} 
\end{figure}

\begin{figure}[h!]
\centering \includegraphics[width=4in]{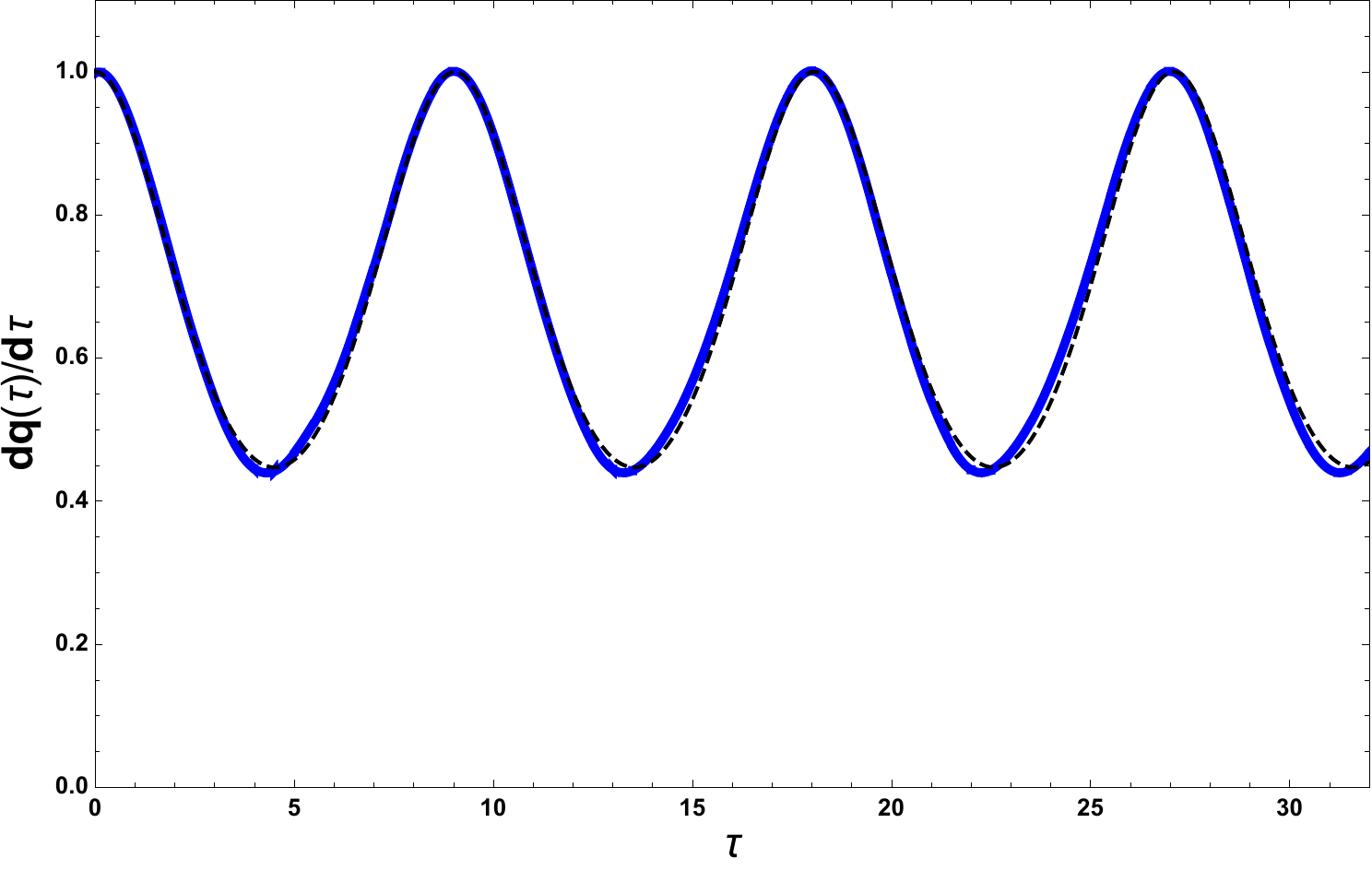}
\caption{Velocity of the ignition interface for problem \eqref{eq:i1}--\eqref{eq:i6} with $\Theta_i=1/2,$ $\Sigma_{min}=0.6 ~(a=0.2)$ and $\eps=0.05$ (blue solid line)
and its asymptotic value obtained from expression \eqref{eq:fr9} (black dashed line).}
\label{f:6} 
\end{figure}

\begin{figure}[h!]
\centering \includegraphics[width=4in]{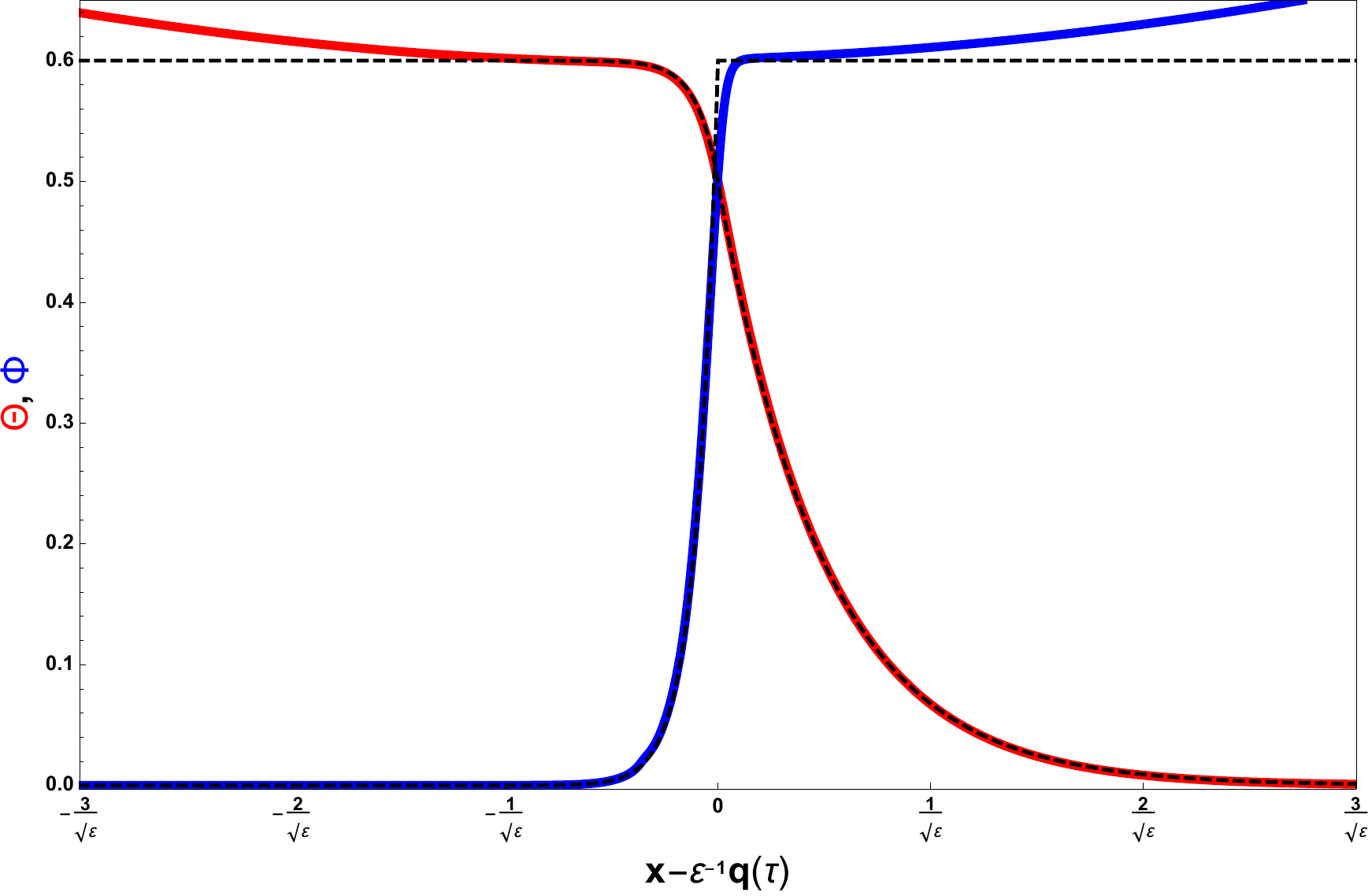}
\caption{A snapshot of the traveling front profile for problem \eqref{eq:i1}--\eqref{eq:i6} with $\Theta_i=1/2,$ $\Sigma_{min}=0.6 ~(a=0.2)$ and $\eps=0.05$ 
in $1/\sqrt{\eps}$ vicinity of the ignition interface at the instance when the front is located at the local minimum of the initial concentration profile. 
 Red and blue solid lines indicate the profile of the temperature and concentration of the deficient reactant. Black dashed line represent these quantities obtained by the asymptotic formulas \eqref{eq:fr4} and \eqref{eq:fr5}.  }
\label{f:7} 
\end{figure}

\begin{figure}[h!]
\centering \includegraphics[width=4.2in]{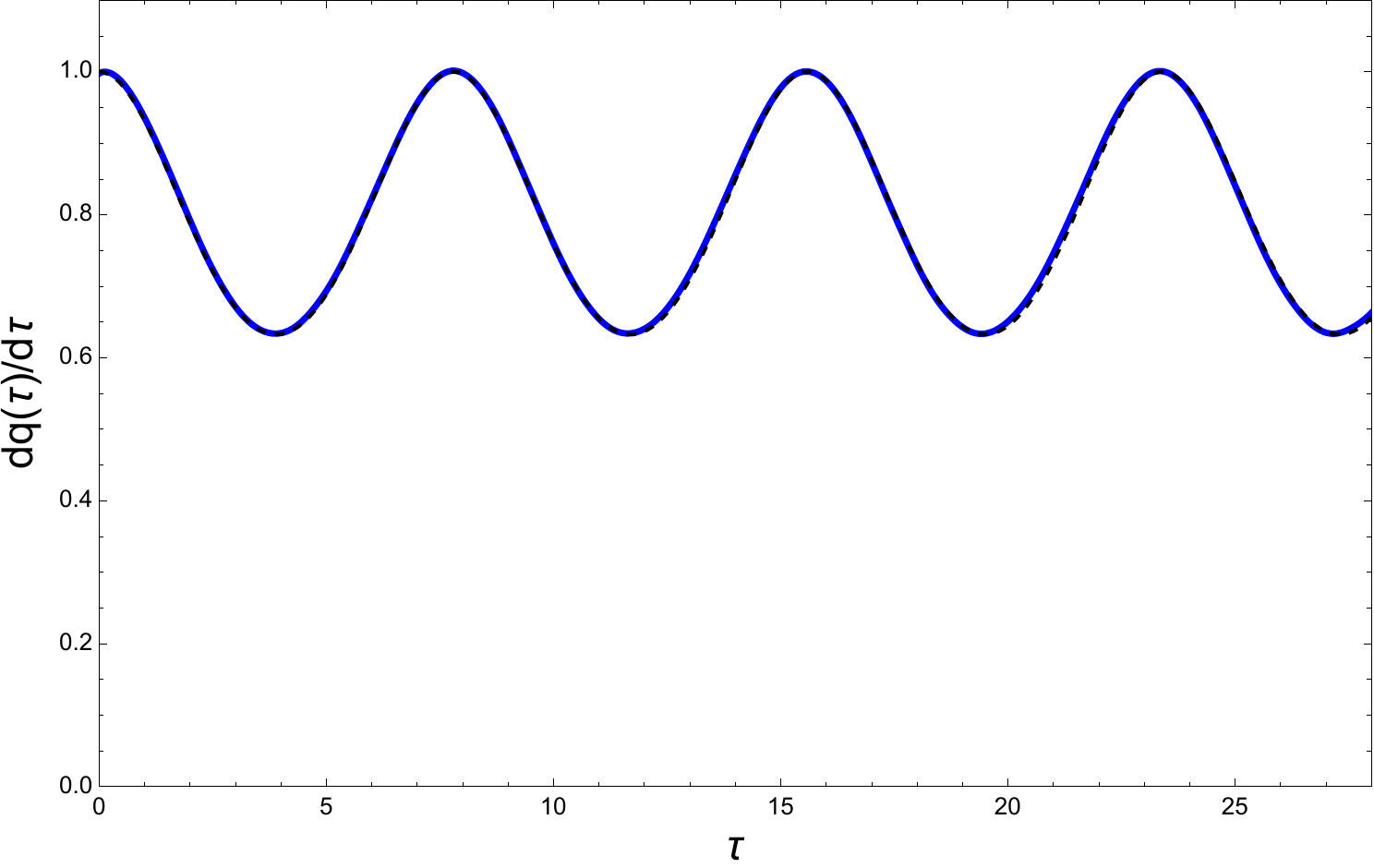}
\caption{Velocity of the ignition interface for problem \eqref{eq:i1}--\eqref{eq:i6} with $\Theta_i=1/2,$ $\Sigma_{min}=0.7 ~(a=0.15)$ and $\eps=0.05$ (blue solid line)
and its asymptotic value obtained from expression \eqref{eq:fr9} (black dashed line).}
 \label{f:8} 
\end{figure}

Let us also note that the  function $dq(\tau)/d\tau$ is a periodic function  with period 
\begin{eqnarray}
T\left(\lambda \right)=4 F\left(\frac{\pi}{2}, \lambda \right),
\end{eqnarray}
where $F\left(\frac{\pi}{2}, \lambda \right)$ is the elliptic integral of the first kind \cite{AS}.
It is easy to check that the period of oscillation for the periodic front  is an increasing function of the parameter  $\lambda$. Moreover, one has
 $T\left(\lambda \right) \to 2\pi$ when $\lambda \to 0$ and $T\left(\lambda \right) \to \infty$ when $\lambda \to 1$.
The dependency of the period on $\lambda$ is depicted in Figure \ref{f:4}.
The comparison of direct numerical simulation of problem  \eqref{eq:i1}-\eqref{eq:i6} with the asymptotic solution constructed in this section shows  excellent agreement in terms of
position, velocity and local structure of the ignition front provided $\Sigma_{min}$ is sufficiently larger than the ignition temperature. 
Figure \ref{f:5} shows several snapshots of the propagating traveling front over one period of oscillation as a solution of problem \eqref{eq:i1}--\eqref{eq:i6} with $\Theta_i=0.5,$
$\Sigma_{min}=0.6~(a=0.2)$ and $\eps=0.05.$
The velocity of the ignition front obtained in this simulation is very close to  the one obtained from the asymptotic expression (the maximal relative error is about $5\%$),
see Figure \ref{f:6}. Moreover, as expected the local structure of the traveling front is quite close to the one given by \eqref{eq:fr4}, \eqref{eq:fr5}  in a $1/\sqrt{\eps}$ vicinity of the ignition interface, see Figure \ref{f:7}.
As clearly seen from Figure \ref{f:5}, outside of the $1/\sqrt{\eps}$ vicinity of the ignition interface, the periodic front exhibits an oscillatory pattern in burned fuel that decays due to thermal diffusion. 
This oscillatory tail does not significantly influence the propagation of the ignition interface.
For smaller values of the parameter $a$ (that is for larger $\Sigma_{min}$) the accuracy of the asymptotic predictions is even better. Figure \ref{f:8} shows the velocity of propagation
of the flame front obtained by means of direct numerical simulation of problem \eqref{eq:i1}--\eqref{eq:i6} and from asymptotic expression \eqref{eq:fr9}  with $\Theta_i=0.5,$ 
$\Sigma_{min}=0.7~(a=0.15)$ and $\eps=0.05$. In this case, the relative error of asymptotic
prediction is below $1.5\%.$ 
 Similar behavior is also seen for other values of the ignition temperature.
However, when the value $\Sigma_{min}$ gets closer to $\Theta_i$ the picture changes substantially
and the behavior of flames  fronts become very different from predictions of asymptotic formulas.
This discrepancy, to no surprise,  is due to the instability of the periodic front
that occurs when $\Sigma_{min}$ gets closer to $\Theta_i$.
 As will be shown in the
next section, the stability of the periodic flame front  is controlled by a parameter
\begin{eqnarray}
\Lambda=\frac{\Theta_i}{\Sigma_{min}}.
\end{eqnarray}
When the value of this parameter is below some critical value $\Lambda^*\approx 6/7\approx 0.857$ the
periodic front appears to be stable but exhibits instabilities when $\Lambda$ exceeds $\Lambda^*$.
Note that the velocity of the periodic flame fronts depicted in Figure \ref{f:8}
correspond to value of $\Lambda=0.714$ and hence represents a stable flame front
well within the stability region. Velocity profile depicted in Figure \ref{f:6} 
corresponds to  $\Lambda=0.833$  that is 
slightly below the stability threshold $\Lambda^*.$ 

The behavior of flame fronts above the stability threshold 
 depends  quite sensitively on specific  values of controlling
parameters $\Theta_i$ and $\Sigma_{min}$.  We will discuss this behavior in Section \ref{s:pattern}.
In the next section we will discuss the linear stability of the periodic flame fronts.

\section{ Linear stability of fronts}\label{s:stab}
As evident from the discussion in the preceding sections, the local behavior of
periodic flame fronts is determined (in the first approximation) by traveling front solutions
of auxiliary problem \eqref{eq:fr1}--\eqref{eq:fr3}. It is then expected that 
the stability of the periodic fronts can be associated with the stability of solutions
of  this problem given by  \eqref{eq:fr4}--\eqref{eq:fr6} with $\Sigma=\Sigma_{min}$.
In what follows we briefly discuss the stability of these solutions as solutions of  problem \eqref{eq:i1}.

\begin{figure}[h]
\centering \includegraphics[width=5in]{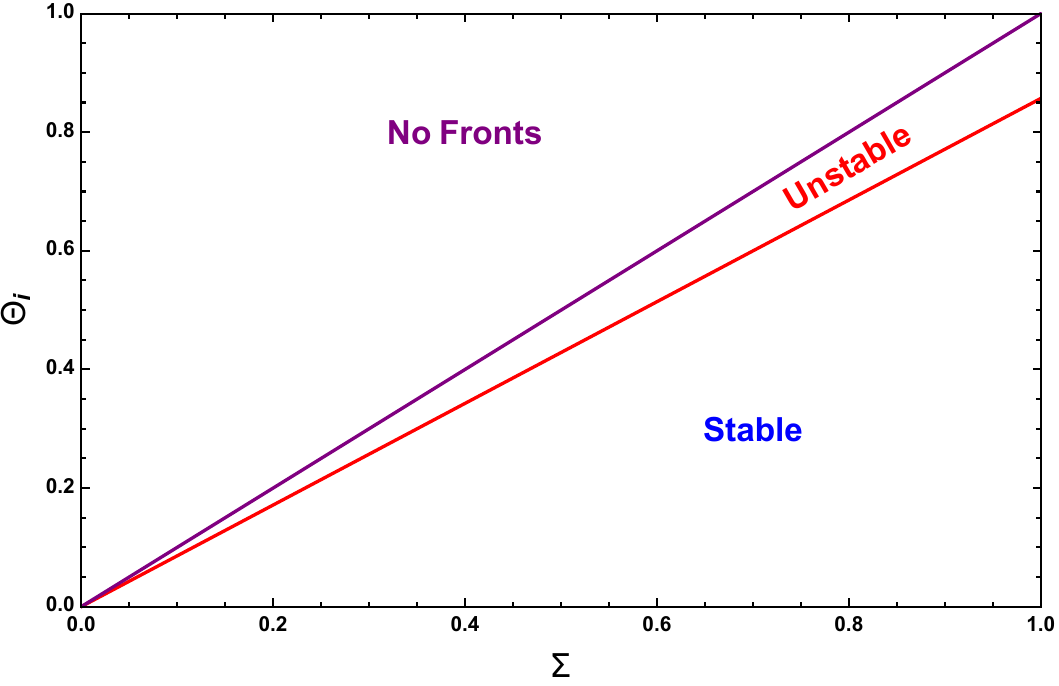}
 \caption{
Neutral stability curve and regions of  stability, instability and non existence of fronts.}
\label{f:9} 
\end{figure}

\begin{figure}[h!]
\centering \includegraphics[width=3.7in]{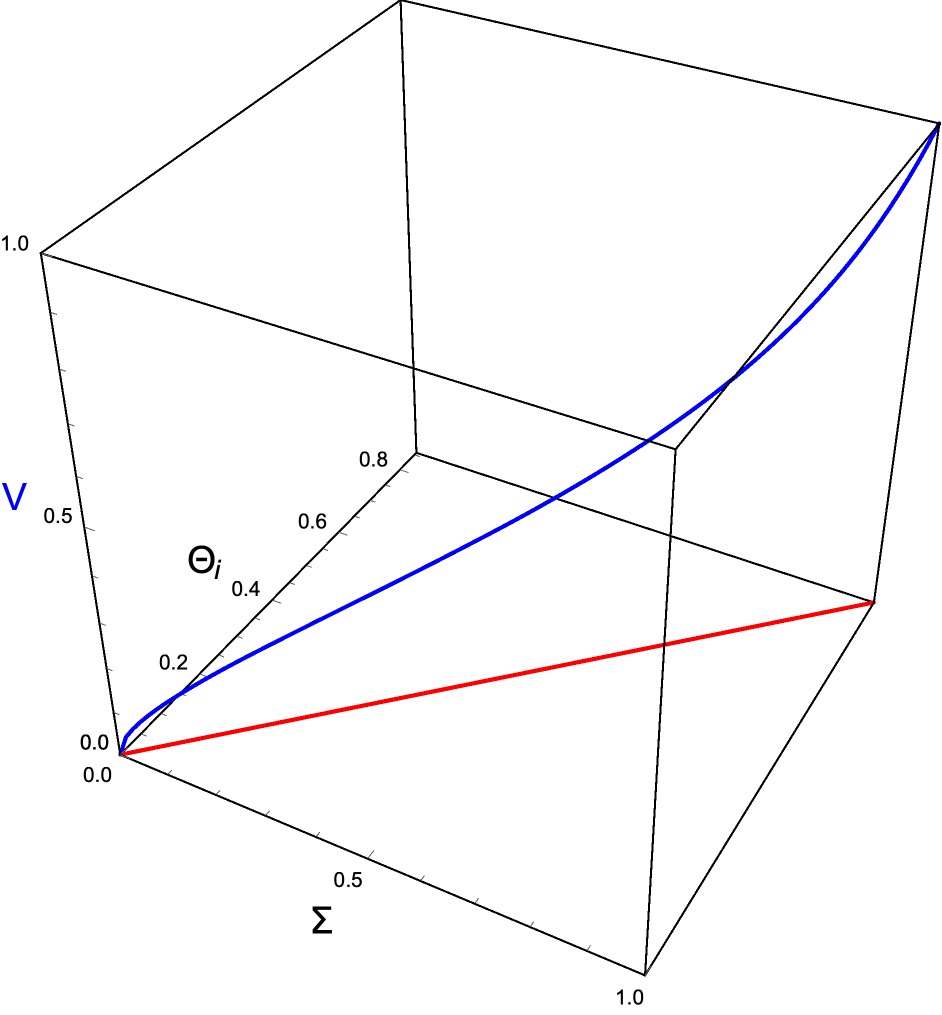}
 \caption{
Velocity (blue line) along the neutral stability curve (red line). }
\label{f:10} 
\end{figure}

\begin{figure}[h!]
\centering \includegraphics[width=3.7in]{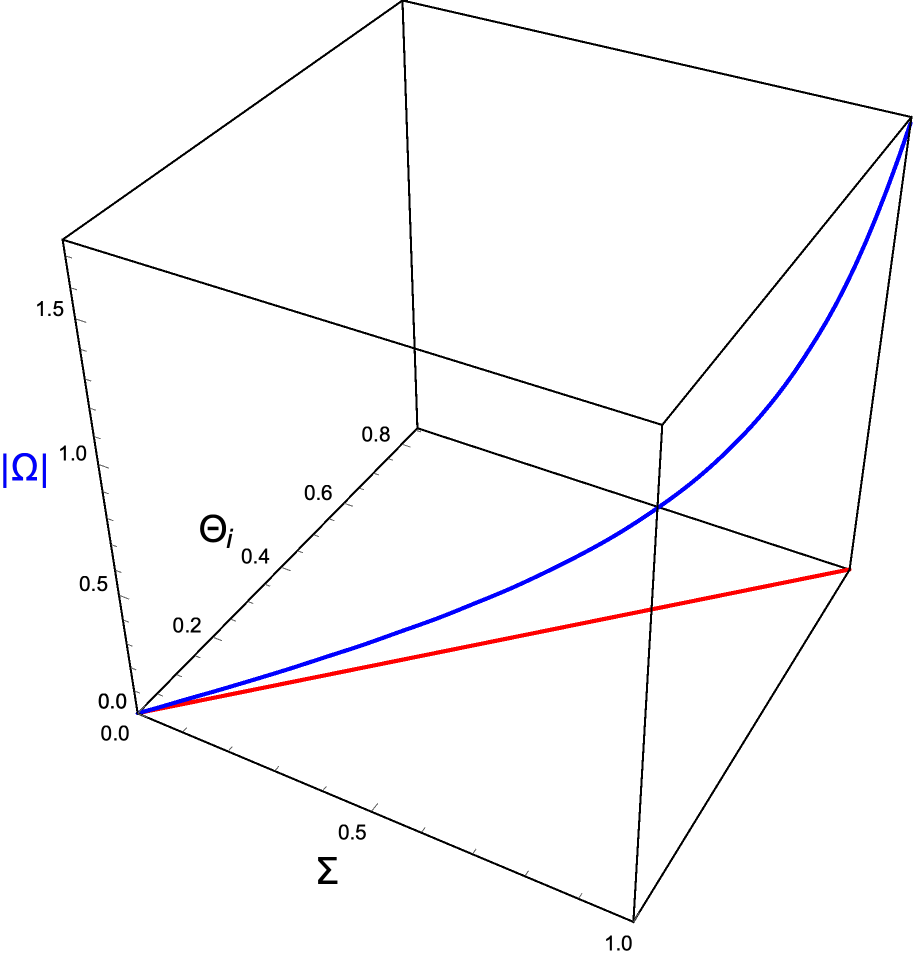}
 \caption{
Frequency (blue line) along neutral stability curve (red line). }
\label{f:11} 
\end{figure}

Setting
\begin{eqnarray}\label{eq:st1}
&&\Theta(x,t)=\theta(\eta)+\exp(\omega t) \vartheta(\eta),\\
&&\Phi(x,t)=\phi (\eta)+\exp(\omega t) \varphi(\eta),\\
&& \eta=x-Vt-\exp(\omega t)v,
\end{eqnarray}
substituting this ansatz into \eqref{eq:i1} and assuming that $\vartheta,\varphi$ and $v$ are infinitesimally small
after some tedious, but straightforward computations similar to these in \cite{BGKS}  one ends up with the following dispersion repletion,
\begin{eqnarray}\label{eq:st2}
{\cal H}(\Theta_i, \Sigma,\omega)=0, \quad  \omega\in\mathbb{C}.
\end{eqnarray}
where
\begin{eqnarray}\label{eq:st3}
&&{\cal H}(\Theta_i, \Sigma,\omega)=\left[ \Theta_i-2V(1-\Theta_i)p\right] \times \nonumber\\
&& \left[A+ \left(\frac{A+\omega}{V}\right)^2 \right]
+\frac{A\Sigma}{V} \left[p-\frac{A+\omega}{V}\right],
\end{eqnarray}
and 
\begin{eqnarray}\label{eq:st4}
p=\frac12 \left[ -V+\sqrt{V^2+4\omega}\right].
\end{eqnarray}
The neutral stability condition is then obtained as a solution of \eqref{eq:st2} restricted to $\omega$ with zero real part that is:
\begin{eqnarray}\label{eq:st5}
{\cal H}(\Theta_i, \Sigma, i\Omega)=0, \quad \Omega\in\mathbb{R}.
\end{eqnarray}
The neutral stability curve $\Theta^*(\Sigma)$ obtained as a solution of equation \eqref{eq:st5} 
turned out to be a straight line  with the accuracy above $0.5\%$
given as 
\begin{eqnarray}
\Theta^*(\Sigma)\approx \Theta^*_i(\Sigma)\approx \Theta_i^*(1) \Sigma \approx 0.857 \Sigma, \quad \Sigma\in(0,1].
\end{eqnarray} 
We also note that $\Theta^*(1)=6/7$ as follows from \cite[Section 4]{CMB1}.
It is also easy to check that for $ \Theta_i^*(\Sigma)<\Sigma \le \Theta_i$ the traveling fronts are  unstable.  For $0<\Theta_i\le \Theta_i^*(\Sigma)$ the traveling fronts are   linearly stable
and for $\Theta_i>\Sigma,$ thanks to \eqref{eq:fr7}, the fronts do not exist, see Figure \ref{f:9}.
The velocity of the front along the neutral stability curve and frequency along the neutral stability curve, obtained as a solution of equation \eqref{eq:st5},
are shown on Figures \ref{f:10} and \ref{f:11}.
Thus, in full agreement with the numerical simulation, we conclude that the instability of
periodic fronts takes place when $\Sigma_{min}$ is below the stability threshold. 
One may  expect that, similar to the case of homogeneous solid fuel concentration, there is a rich family of  instability patterns which get more and
more complex as $\Sigma_{min}$ decreases. In the following section we 
will discuss one of (apparently many possible) regimes of propagation:
propagation - extinction- diffusion - reignition regime. This regime  occurs exclusively
due to the presence of inhomogeneity in solid reactant and ,to the best of our knowledge, was not reported in the literature.

\section{Propagation - extinction - diffusion - reignition regime}\label{s:pattern}

In this section we will discuss the behavior of the combustion fronts above the stability threshold.
This behavior is rather sensitive to the specific values of the controlling parameters $\Theta_i$ and
$\Sigma_{min}$. When  $\Theta_i$ is relatively large
$\Theta_i>\Theta_i^*\approx 0.37$ the crossing of the stability threshold leads to extinction.
\begin{figure}[h!]
\centering \includegraphics[width=4in]{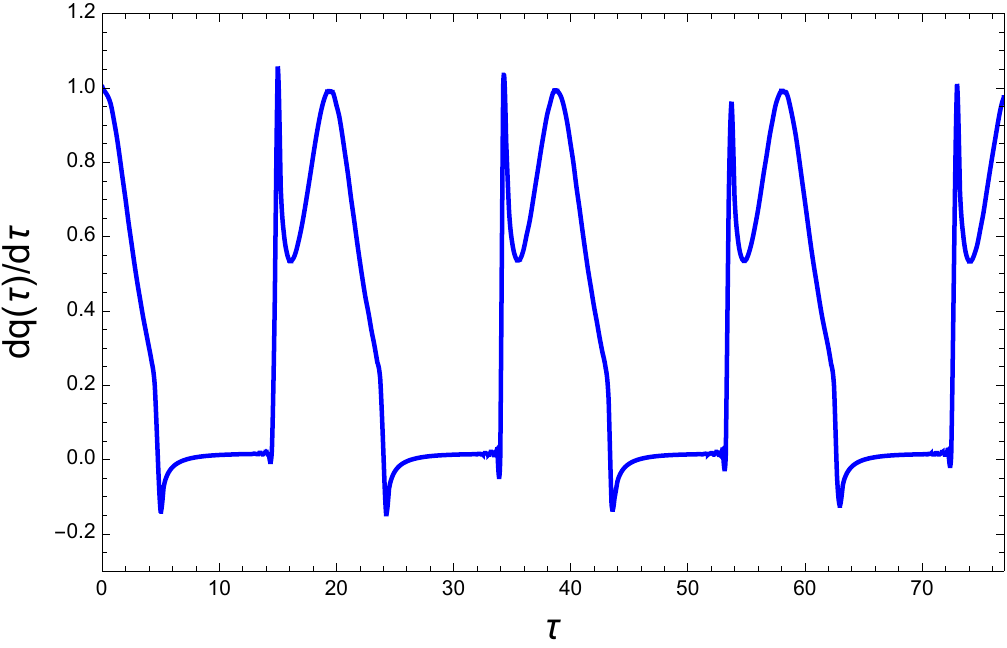}
 \caption{
The velocity of the ignition interface obtained form solution of problem \eqref{eq:i1}--\eqref{eq:i6}
with $\Theta_i=0.3,$ $\Sigma_{min}=0.32 ~(a=0.34)$ and $\eps=0.05$.  }
\label{f:12} 
\end{figure}
However, when $\Theta_i$ is smaller than $\Theta^*$ and $\Lambda$ is slightly above the stability threshold $\Lambda^*,$
an interesting  propagation - extinction - diffusion - reignition regime emerges.
\begin{figure}[h!]
\centering \includegraphics[width=3.2in]{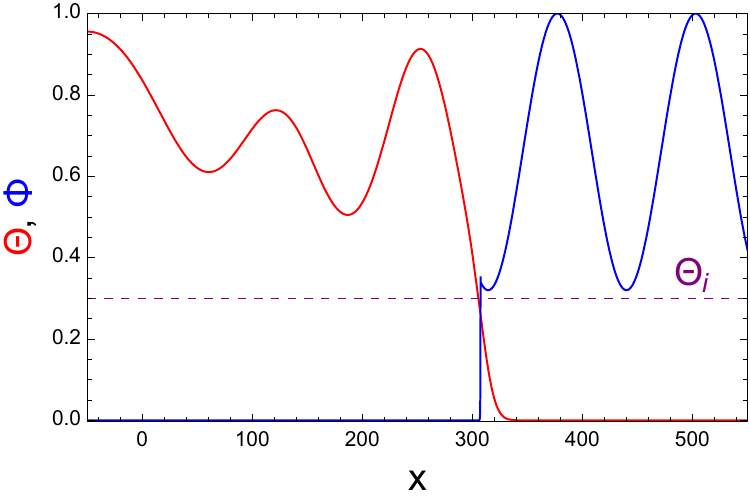}
\centering \includegraphics[width=3.2in]{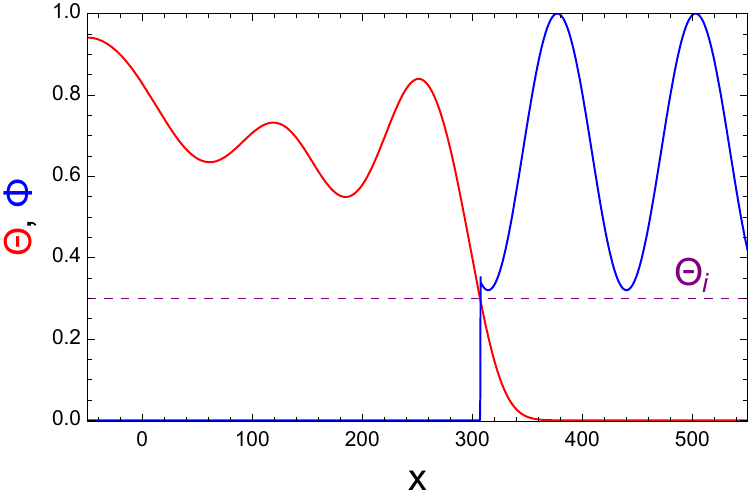}\\
\centering \includegraphics[width=3.2in]{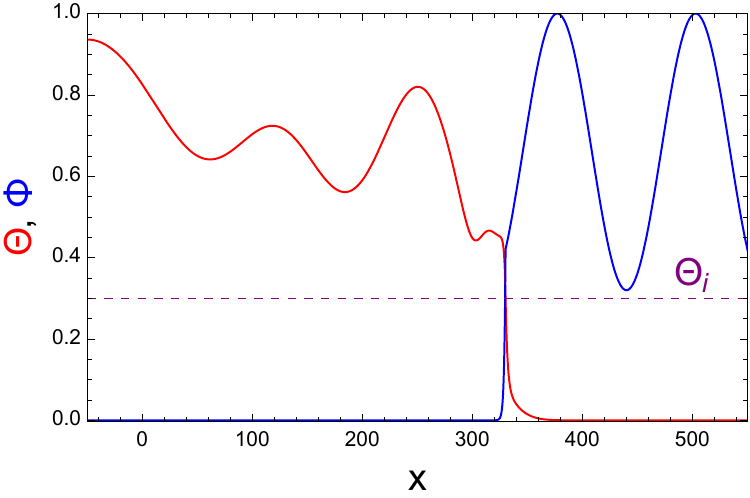}
\centering \includegraphics[width=3.2in]{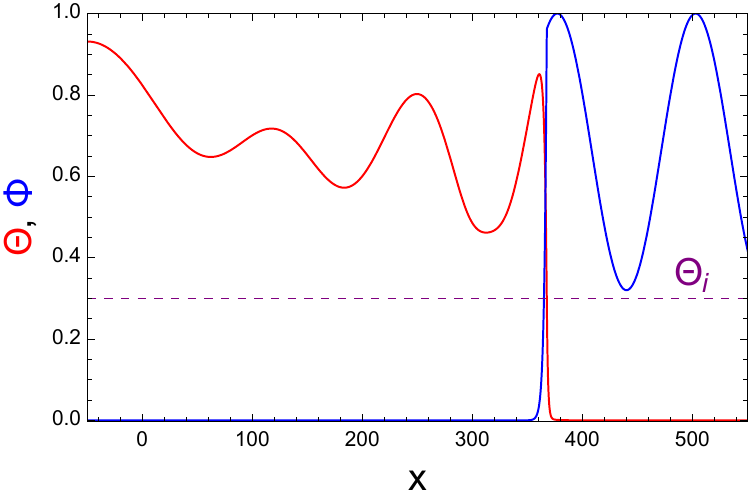}\\
\centering \includegraphics[width=3.2in]{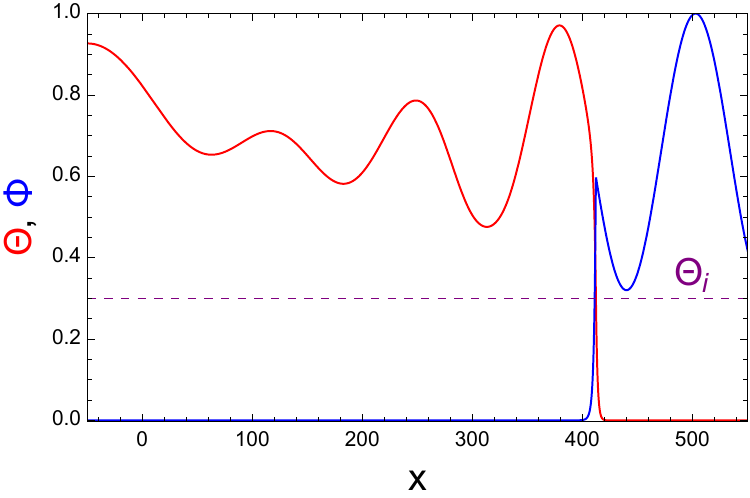}
\centering \includegraphics[width=3.2in]{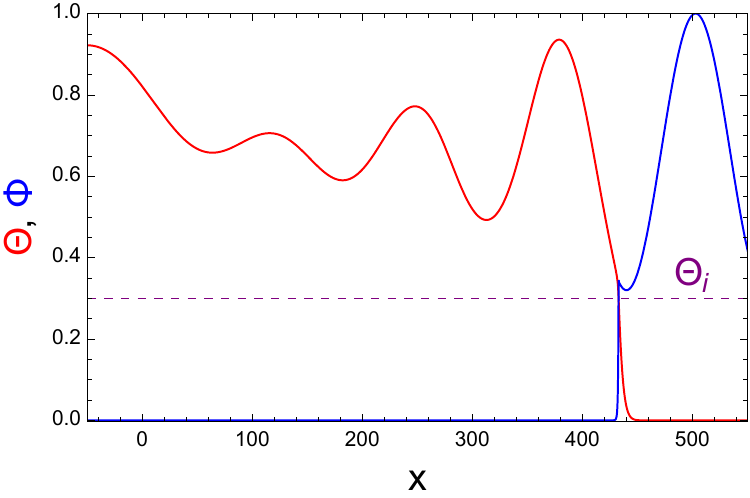}
\caption{Representative snapshots of flame  front profile (rad)  and concentration profile (blue) for problem \eqref{eq:i1}--\eqref{eq:i6} with  $\Theta_i=0.3$ and $\Sigma_{min}=0.32 ~(a=0.34)$ and $\eps=0.05$.
 }
\label{f:13} 
\end{figure}
In this regime comparable parts of the initial concentration curve are above and below the
stability threshold. The dynamics of this regime can be described as follows.
Initially the ignition front propagates as a stable periodic front in the region when concentration
of the deficient reactant is high. As the front  propagates, the concentration decreases and the front
enters in the fuel lean instability region which eventually leads to complete extinction. Theses processes 
proceed at a relatively fast reactive time scale.
 After the extinction takes place, the only mechanism that impacts the flame front
is the diffusion of temperature. This slow process puts dynamics on a characteristic diffusive  time scale.
 The flame front starts to diffuse  and widens. This results in a slow temperature
 increase in the region where concentration of the solid reactant is higher. This dynamic finally results in re-ignition of the reactant that triggers a local explosion which reignites the flame. The flame then starts to propagate  up to the point when the low concentration region is reached. This process
 repeats itself in an  approximately periodic manner. 
 
 Figure \ref{f:12} shows the velocity of the ignition interface when $\Theta_i=0.3$ and $\Sigma_{min}=0.32 ~(a=0.34)$. Figure \ref{f:13} depicts snapshots of
 dynamic of the flame front in this regime. The further increase of oscillations in the initial
 concentration leads to extinction. For $\Theta_i=0.3,$ for example, when $a$ exceeds value
 $0.413$ ($\Sigma_{min}=0.174$) the extinction takes place.

Let us also note that, as evident from Figure \ref{f:12}, the flame front in propagation - extinction - diffusion - reignition regime exhibits propagation with negative velocity. This type of behavior 
was reported for a model of nearly equi-diffusive combustion \cite{add1}, but known to be absent in
gasless combustion of homogeneous samples \cite{add2}.

 \bigskip
 
\noindent {\bf Acknowledgments.}  The work of AM, LK, GS and PVG was supported, in part, by the US-Israel Bination Science Foundation (Grant 2020-005).

\bigskip

\noindent{\bf Declaration of Competing Interest.} Authors declare no competing interests.

\end{document}